\documentclass[%
 aip,
 apl,
 amsmath,amssymb,
 reprint,%
]{revtex4-1}

\usepackage{CJK}
\usepackage{graphicx}
\usepackage{dcolumn}
\usepackage{bm}
\usepackage{siunitx}
\DeclareSIUnit\sq{\ensuremath{\Box}}
\begin{document}
\begin{CJK}{UTF8}{}
\preprint{AIP/123-QED}

\title[Increased multiplexing of superconducting microresonator arrays by post-characterization adaptation of the on-chip capacitors]{Increased multiplexing of superconducting microresonator arrays by post-characterization adaptation of the on-chip capacitors}

\author{S. Shu (\CJKfamily{gbsn}{舒诗博})}
	\email{shu@iram.fr}
	\affiliation{ 
Institut de RadioAstronomie Millim\'{e}trique, 38406 Saint Martin d'H\`{e}res, France
}

\author{M. Calvo}
	\affiliation{
Institut N\'{e}el, CNRS and Universit\'{e} Grenoble Alpes, 38042 Grenoble, France
}

\author{J. Goupy}
	\affiliation{
Institut N\'{e}el, CNRS and Universit\'{e} Grenoble Alpes, 38042 Grenoble, France
}

\author{S. Leclercq}
	\affiliation{ 
Institut de RadioAstronomie Millim\'{e}trique, 38406 Saint Martin d'H\`{e}res, France
}

\author{A. Catalano}
\affiliation{
Institut N\'{e}el, CNRS and Universit\'{e} Grenoble Alpes, 38042 Grenoble, France
}
\affiliation{
LPSC, CNRS and Universit\'{e} Grenoble Alpes, 38026 Grenoble, France
}

\author{A. Bideaud}
\affiliation{
Institut N\'{e}el, CNRS and Universit\'{e} Grenoble Alpes, 38042 Grenoble, France
}

\author{\\A. Monfardini}
\affiliation{
Institut N\'{e}el, CNRS and Universit\'{e} Grenoble Alpes, 38042 Grenoble, France
}

\author{E.F.C. Driessen}
	\affiliation{ 
Institut de RadioAstronomie Millim\'{e}trique, 38406 Saint Martin d'H\`{e}res, France
}

\date{\today}

\begin{abstract}
We present an interdigitated capacitor trimming technique for fine-tuning the resonance frequency of superconducting microresonators and increasing the multiplexing factor. We first measure the optical response of the array with a beam mapping system to link all resonances to their physical resonators. Then a new set of resonance frequencies with uniform spacing and higher multiplexing factor is designed. We use simulations to deduce the lengths that we should trim from the capacitor fingers in order to shift the resonances to the desired frequencies. The sample is then modified using contact lithography and re-measured using the same setup. We demonstrate this technique on a 112-pixel aluminum lumped-element kinetic-inductance detector array. Before trimming, the resonance frequency deviation of this array is investigated. The variation of the inductor width plays the main role for the deviation. After trimming, the mean fractional frequency error for identified resonators is \num{-6.4e-4}, with a standard deviation of \num{1.8e-4}. The final optical yield is increased from 70.5\% to 96.7\% with no observable crosstalk beyond $\SI{-15}{dB}$ during mapping. This technique could be applied to other photon-sensitive superconducting microresonator arrays for increasing the yield and multiplexing factor.

\end{abstract}


\maketitle
\end{CJK}

Superconducting microresonators are of great interest for many applications like photon detection~\cite{Day:2003a}, readout of superconducting quantum interference devices~\cite{Irwin:2004a} and quantum computation~\cite{Wallraff:2004a}, thanks to their high sensitivity, simple structures and intrinsic frequency-domain multiplexing properties~\cite{Zmuidzinas:2012a}. In the photon detection field, superconducting microresonator based microwave kinetic inductance detectors (MKID)~\cite{Day:2003a,Zmuidzinas:2012a,Baselmans:2012b} have been widely developed for astronomical observations from millimeter waves to X-rays~\cite{Mazin:2013a,Baselmans:2017a,Monfardini:2010a,Adam:2018a,Ulbricht:2015a}. MKID is a non-equilibrium superconducting detector, utilizing the kinetic inductance of superconductors. When an incident photon has energy larger than the superconducting gap ($h\nu> 2\Delta$), it can break Cooper pairs and create quasi-particles. The reduction of the number of Cooper pairs changes the kinetic inductance of the superconducting film and the resulting changes of resonance amplitude and phase response can be read out through coaxial cables. Based on the intrinsic frequency multiplexing property, hundreds of resonators can be read out simultaneously, through a single feedline. Instead of using distributed capacitors and inductors, the lumped-element kinetic-inductance detector (LEKID)~\cite{Doyle:2008a}, consists of a lumped-element inductor and capacitor. The inductor part is designed to absorb the incident photons and the resonance frequency is tuned by changing the interdigital capacitor (IDC). This leads to more flexible pixel designs in the photon detection field.

\begin{figure}
\includegraphics[width=3.2in]{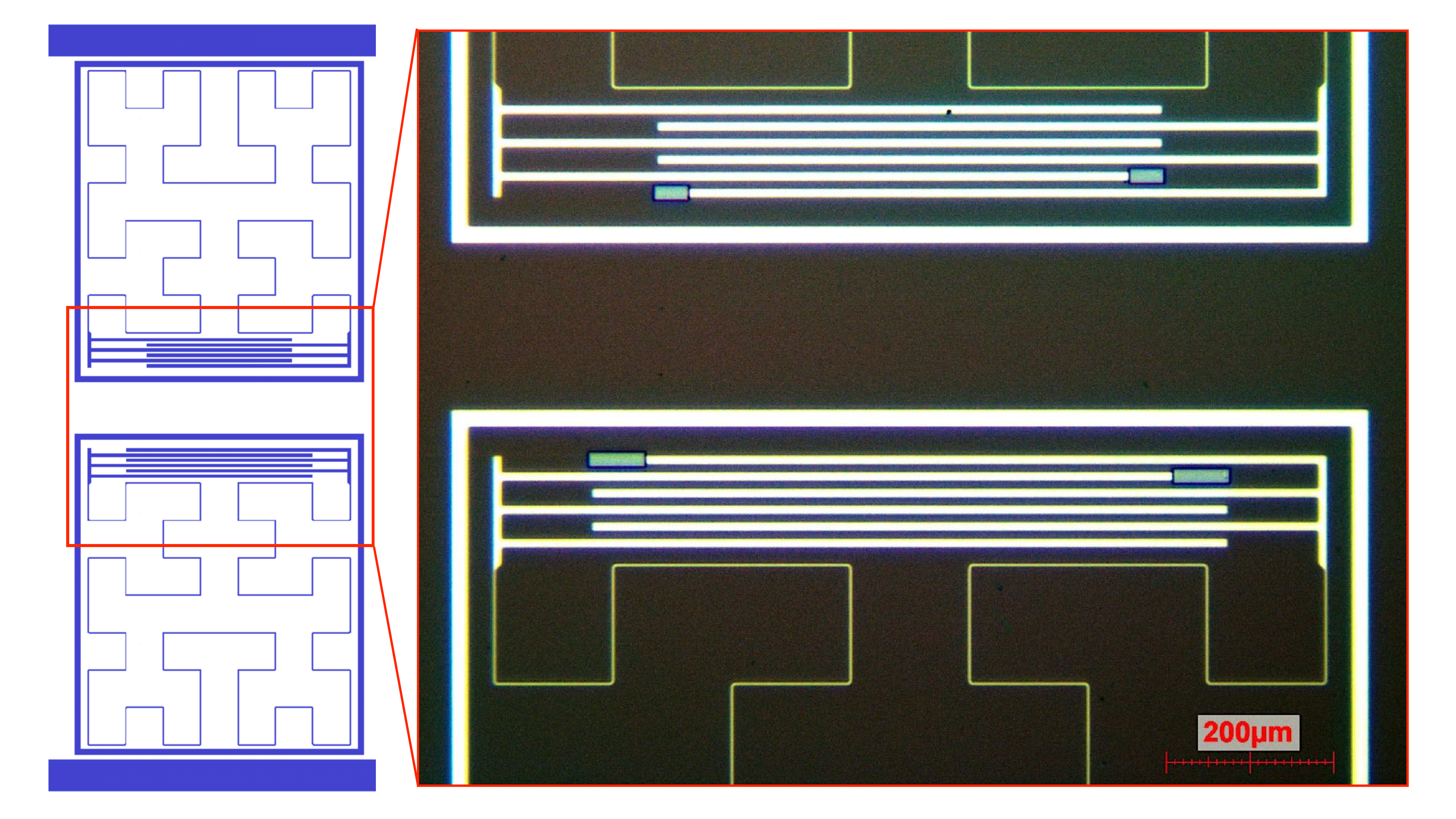}
\caption{\label{fig:kid} (Left) Schematic drawing of two MKID resonators. The resonance frequencies are tuned by the IDC finger lengths of different resonators. Inside a resonator all IDC fingers have the same length before trimming. (Right) The actual array covered with resist after the final etching step of the IDC trimming process. Only the outside pairs of IDC fingers were trimmed (less bright rectangles). The trimmed lengths are calculated based on the needed frequency shifts.}
\end{figure}
\begin{figure*}
\includegraphics[width=0.97\textwidth,trim={0cm 8.9cm 0.1cm 8.2cm},clip]{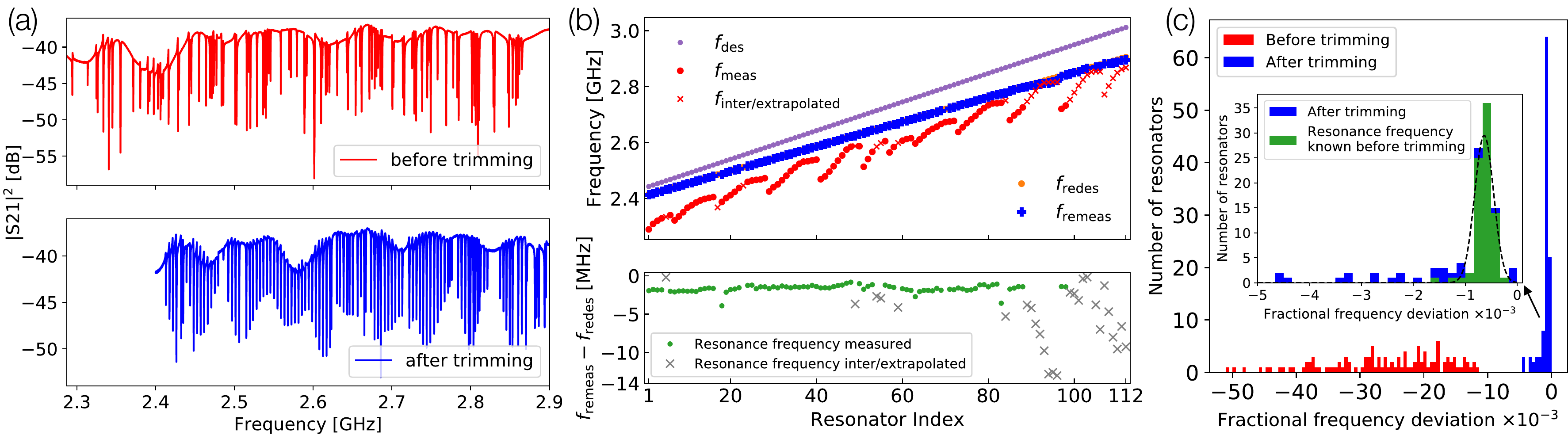}
\caption{\label{fig:fdiff} (a) Comparison of VNA measured feedline transmission S21 for the MKID array before (red) and after (blue) the trimming process. (b) Design ($f_{\textrm{des}}$, purple dot) and measured ($f_{\textrm{meas}}$, red dot) resonance frequencies before trimming, determined using the optical mapping system (upper panel). The resonance frequencies of unidentified resonators ($f_{\textrm{inter/extrapolated}}$, red cross) are inter/extrapolated based on the smooth variation of the frequency deviation (Fig.~\ref{fig:fmap}(a)). The re-designed resonance frequencies ($f_{\textrm{redes}}$, orange dot) for trimming and the re-measured resonance frequencies ($f_{\textrm{remeas}}$, blue dot) after trimming are plotted for comparison. The bottom panel shows the absolute frequency deviation between $f_{\textrm{remeas}}$ and $f_{\textrm{redes}}$. The resonators with $f_{\textrm{meas}}$ (green dots) have a mean value of \SI{1.7}{MHz} and a standard deviation of \SI{0.46}{MHz}. The resonators with $f_{\textrm{inter/extrapolated}}$ (gray crosses) show a larger deviation. (c) Histogram of the fractional resonance frequency deviation $\delta f/f$ before (red) and after (blue) the trimming process. After trimming the deviation is dramatically decreased, even for the inter/extrapolated resonators. In the inset, the green histogram shows the distribution of the resonators of which the exact frequency shift was known. The blue points show the resonators that were estimated using the inter/extrapolation described in the text. A Gaussian fit of the resonance frequencies of the identified resonators is plotted (dashed line), which gives a mean value of $\mu=\num{-6.4e-4}$ and a standard deviation of $\sigma =\num{1.8e-4}$.}
\end{figure*}

As the multiplexing density increases, the non-uniformity of frequency spacing causes crosstalk when two resonances have a frequency collision. Compared with the design frequency, the measured resonance frequency can shift by tens of MHz. Usually the bandwidth of a feedline is limited, for example to 500 MHz for the NIKA2 instrument~\cite{Adam:2018a}, and the frequency spacing between adjacent resonators is designed to be several MHz. The minimal spacing is limited by the width of the resonance, which is determined by the quality factor $Q$. For ground-based millimeter-wave continuum detection, the $Q$ is limited by the high sky background load and is around \num{8e3} for NIKA2. Since the measured resonance frequency has the same or even larger shift compared with the spacing, the crosstalk caused by the frequency collisions happens frequently and decreases the number of functional pixels. For photon detection application, currently this is the main limitation on MKID array yield~\cite{Calvo:2016a,Mazin:2013a}.

There are various factors causing the intrinsic resonance frequency deviation, such as variation in the film properties~\cite{Vissers:2013b,Adane-Jr:2016a}, fabrication process inhomogeneity~\cite{McKenney:2018a} and resonator design~\cite{Noroozian:2012a,Yates:2014a}. For titanium-nitride films the sheet resistance, superconducting critical temperature ($T_c$) and capacitor etching depth~\cite{Vissers:2013b,McKenney:2018a} are the dominant factors. For thin Al films usually the film thickness variation is the dominant factor~\cite{Adane-Jr:2016a}, however, we find that for our test array the variation in the inductor width is dominant. To decrease the deviations, several methods have been previously explored~\cite{Vissers:2013b,Noroozian:2012a,Vissers:2013a}. By performing post-characterization adaption, we do not necessarily need to know the causes of the frequency deviation. ~\citet{Liu:2017a} \cite{Liu:2017b} have demonstrated a capacitor trimming technique to tune the resonance frequency accurately. They have used a LED mapper, deployed on the same stage as the sample inside their cryostat, to find the physical resonators corresponding to the measured resonances~\cite{Liu:2017a}. This LED mapper should be fabricated depending on the number and design of the array and cannot be easily used for a different array. Also the linear relation of capacitance and IDC finger length cannot be extended to quasi-lumped element IDC. In this Letter, we present a
capacitor trimming technique for quasi-lumped element KIDs using a beam mapping system. Compared with the LED mapper, the beam mapping measurement is a necessary characterization for photon detectors and no extra parts or wires are needed inside the cryostat. Also this mapping system is universal for all detector arrays without limitation on pixel number or design. We also demonstrate a method to tune quasi-lumped element IDC based on electromagnetic simulation. Using a NIKA2 test array, the comparison before and after trimming is discussed including the optical properties of the array.

We applied this technique on a prototype NIKA2 LEKID array~\cite{Shu:2018a}, shown in Fig.~\ref{fig:kid}, which was fabricated from a \SI{21.6}{nm} thick Al film on a \SI{250}{\um} thick, high-resistivity silicon substrate. To get dual-polarization sensitivity the inductors are designed in a third-order Hilbert curve~\cite{Roesch:2012a}. The capacitors are designed with $6$ IDC fingers. The Al film properties are independently measured and give a thickness of \SI{21.6}{\nm}, a transition temperature of \SI{1.46}{\K} and a sheet resistance at \SI{4.2}{K} of $\SI{1.14}{\ohm\per\sq}$. The sheet inductance of $\SI{1.09}{pH\per\sq}$ is then estimated from Mattis-Bardeen theory~\cite{Mattis:1958a,BARENDS:2009c}. A \SI{200}{\nm} Al film was deposited on the backside of the wafer as a backshort to maximize the absorption in the \SI{260}{\GHz} band and serves, at the same time, as the ground plane of the microstrip readout line. With the measured sheet inductance of $\SI{1.09}{pH\per\sq}$, the resonance frequency range $f_{\textrm{des}}$ is expected from \SIrange{2.379}{2.941}{\GHz} with \SI{5.06}{\MHz} frequency spacing. After the first fabrication the sample was cooled down to \SI{80}{\milli\K} in a $^3\textrm{He}$-$^4\textrm{He}$ dilution refrigerator with an optical access. The transmitted optical band is defined by a \SI{169}{\GHz} high pass and \SI{304}{\GHz} low pass filters, but the detector is only sensitive in the range from \SIrange{230}{270}{\GHz}, as defined by the backshort distance.

\begin{figure*}
\includegraphics[width=0.97\textwidth,trim={0.4cm 9.1cm 0cm 8.3cm},clip]{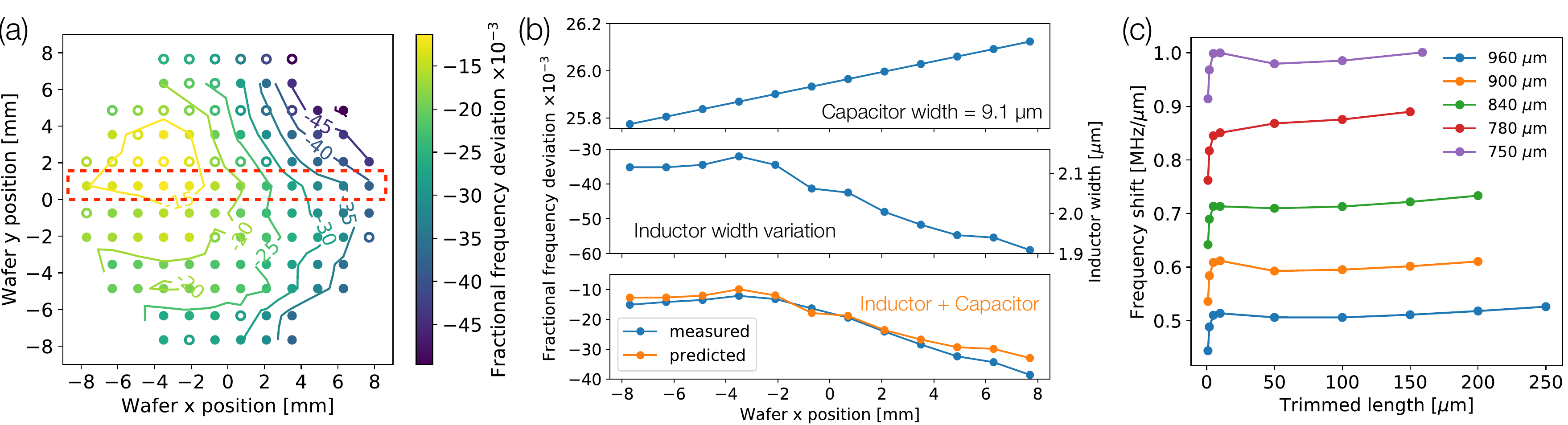}
\caption{\label{fig:fmap} (a) Mapping of the fractional frequency deviation $\delta f/f$ with respect to the position of resonator on wafer. The resonance frequencies of unidentified resonators (unfilled circle) are inter/extrapolated using radial basis functions. (b) For selected resonators, shown in the red rectangle in (a), the fractional frequency deviations due to the decreased IDC finger width, estimated from simulation, and induced by the variation of the inductor width, measured with SEM, are plotted in the upper and middle panel respectively. The bottom panel shows the sum of both deviations from capacitor and inductor, consistent with the measured one. The maximum deviation between estimation and measurement is smaller than \num{6e-3}. (c) Resonance frequency shifts per micrometer trimmed, simulated for different trimmed lengths. Different curves indicate different IDC fingers lengths. Given the IDC finger length and the required frequency shift of a resonator, the length of IDC to be trimmed is derived from this plot.}
\end{figure*}

In the first measurement, the feedline transmission S$21$, shown in Fig.~\ref{fig:fdiff}(a), was measured using a vector network analyzer (VNA) with a mirror placed in front of the cryostat window, which gives an optical loading on the detectors equivalent to a $\SI{10}{\K}$ black body. The measured resonance frequencies $f_{\textrm{meas}}$ range from \SIrange{2.289}{2.870}{\GHz} and the central frequency is shifted by \SI{80.5}{\MHz} from the initial design $f_{\textrm{des}}$. This \SI{3}{\percent} frequency deviation cannot be explained by the Al film thickness variation between \SI{21.5}{nm} and \SI{21.7}{nm}, which was independently measured by ellipsometry, and would correspond to a maximum \SI{13}{MHz} deviation, nor by the resonator-to-resonator electrical crosstalk, which is estimated by simulation to be smaller than \SI{20}{\kHz}~\cite{Shu:2018a}.

The resonance frequencies are linked to the physical resonators using a beam mapping system \cite{Shu:2018a}. The detailed optical setup is shown in the supplementary material. The feedline transmission and the optical performances of the MKIDs are measured in the same cool-down. The beam mapping system consists of a movable \SI{300}{\K} small metal ball in front of a \SI{50}{\K} cold background. The responses of the MKIDs and the position of the metal ball are read out by the NIKEL system~\cite{Bourrion:2016a}. By pinning out the position of the maximum of the response of each MKID, the resonance frequencies are mapped to the physical position of the resonators. In this single measurement, only $79$ out of $112$ resonances were identified with physical resonators, because of the 500 MHz readout bandwidth limitation and crosstalk. We observed a smooth variation of the normalized fractional frequency deviation over the surface of the array (Fig.~\ref{fig:fmap}(a)).

We have measured the capacitor and inductor widths of a selected line of resonators, shown in Fig.~\ref{fig:fmap}(b), using a scanning electron microscope (SEM). These measurements show that the capacitor widths keep a constant value of \SI{9.1}{\um} instead of the designed \SI{10}{\um}, however, the inductor widths $w$ vary from \SIrange{1.9}{2.14}{\um} instead of \SI{2.5}{\um} in the design. The fractional frequency deviation resulting from the \SI{0.9}{\um} narrowed IDC fingers, shown in Fig.~\ref{fig:fmap}(b), are estimated using simulation. The small variation, as a function of frequency (position on wafer), is caused by the quasi-lumped element property. To understand the effect of the variation of the inductor width, we assume the capacitance $C$ is constant and the frequency deviation is derived as
\begin{equation}
\frac{\delta f}{f} = \frac{1}{4}\frac{\delta w}{w},
\label{eq:one}
\end{equation}
where $f= 1/(2\pi\sqrt{LC})$ and the inductance $L\propto 1/\sqrt{w}$, a good estimation for a narrow lumped-element inductor~\cite{Tolpygo:2015a}. Using Eqn.~\ref{eq:one}, the fractional frequency deviation caused by the variation of the inductor width is calculated and shown in Fig.~\ref{fig:fmap}(b). The fractional frequency deviations obtained by adding these two deviations together, is consistent with the measurement and the small difference can be explained by the variation of the inductor width inside one resonator. In principle, this information could be used to adapt the resonance frequencies without the need of an optical measurement, to an accuracy of fractional frequency of \num{6e-3} from Fig.~\ref{fig:fmap}(b). We notice that the resonance frequency is very sensitive to the inductor width, around \SI{250}{\MHz\per\um} at \SI{2.5}{\GHz}. This variation not only changes the resonance frequency but also the sensitivity, which is determined by the quasi-particle density and inductor volume. For low-resistivity Al LEKIDs, to achieve optimal optical coupling, the inductor track is inevitably designed to be narrow with respect to the pixel size and wavelength. Therefore, to decrease the resonance frequency deviation, it is important to improve the Al pattern uniformity during the lithography, development and etching processes. We can also decrease the deviation using the trimming technique, when the physical resonators are characterized.

The trimming process is performed as follows. For the resonances unidentified during the optical mapping, a inter/extrapolation method based on radial basis functions was used to  estimate their frequencies. This is a simple estimation and the inter/extrapolated resonance frequencies are referred as $f_{\textrm{inter/extrapolated}}$. We did not compare the $f_{\textrm{inter/extrapolated}}$ to the resonance frequencies measured by VNA and this estimation can be improved in the future. Alternatively, the missing $33$ resonances could be identified in a second optical run.

The second step is to find the relation between the resonance frequency shift and the trimming lengths. Since our MKIDs have a compact design, the total capacitance cannot be estimated accurately from the total IDC finger length with a linear relationship. We used Sonnet Suites~\cite{Sonnet} to simulate the frequency shifts with different trimmed lengths at selected resonance frequency configurations (Fig.~\ref{fig:fmap}(c)). To increase the precision of length definition during patterning, we only trim one pair of the IDC fingers. The alignment error of the contact lithography along the IDC finger direction is compensated by the symmetry of trimming, compared to trimming only a single IDC finger. Therefore, the total trimmed length on one resonator can be kept the same even with a few \si{\um} alignment error. The simulations were done with an arbitrary sheet inductance \SI{2}{pH\per\sq}. Since the capacitance is not affected by the sheet inductance, the fractional frequency shift caused by trimming is constant for any sheet inductance. Therefore, we could apply the simulation results to our sample without knowing the actual sheet inductance in advance.

Next, we determine the actual trimming length. By trimming the IDC finger shorter, the resonance frequency can only be shifted to a higher frequency. We keep the same frequency order as in the initial design. The largest fractional frequency deviation sets the lower limit of the re-designed resonance frequencies. Given the re-designed frequencies $f_{\textrm{redes}}$ from \SIrange{2.415}{2.905}{\GHz} with spacing \SI{4.4}{\MHz}, the needed fractional frequency shifts are calculated by $(f_{\textrm{redes}}-f_{\textrm{meas}})/f_{\textrm{meas}}$. Then the trimming lengths are determined by a cubic interpolation of the simulation results from the needed fractional frequency shifts, calculated based on Fig.~\ref{fig:fmap}(c).

Finally, the trimming was done using a conventional contact lithography patterning, followed by a wet etching process to trim the IDC fingers, as shown in Fig.~\ref{fig:kid}. Considering the \SI{365}{\nm} exposure wavelength and the alignment accuracy, the length definition accuracy is about \SI{0.5}{\um} and the trimmed frequency accuracy is estimated to be \SI{0.4}{\MHz} on average.

After the trimming process, the sample was cooled down again and the feedline transmission S$21$ was measured with the same setup used before trimming (see Fig.~\ref{fig:fdiff}(a)). The resonances that were identified in the first characterization run were tuned with a standard deviation of $f_{\textrm{remeas}}-f_{\textrm{redes}}$ of only $\SI{0.46}{\MHz}$ after trimming, which is consistent with our fabrication accuracy. We notice that the $f_{\textrm{remeas}}$ after trimming has a \SI{1.7}{\MHz} offset to the $f_{\textrm{redes}}$, an effect that is also observed in other experiments~\cite{Liu:2017b}. Since all resonators were trimmed and no resonator was left as reference, we cannot determine whether this offset is caused by the trimming process or by a change of the film properties. The fractional frequency deviation $(f_{\textrm{remeas}}-f_{\textrm{redes}})/f_{\textrm{redes}}$ of the resonators identified before trimming has a mean value of \num{-6.4e-4} and standard deviation of $\sigma =\num{1.8e-4}$ (Fig.~\ref{fig:fdiff}(c)). This corresponds to a factor of $3$ improvement in terms of fractional frequency deviation compared with previous results~\cite{Liu:2017b}. We also notice that the absolute frequency deviation after trimming is constant and does not increase with increasing resonance frequency in our \SI{490}{\MHz} bandwidth. This indicates that the $\sigma$ of absolute frequency deviation stays constant and the fractional frequency deviation can be extremely small for high frequency resonances. For the inter/extrapolated resonances, the standard deviation of $f_{\textrm{remeas}}-f_{\textrm{redes}}$ is $\SI{3.7}{\MHz}$, because of the inaccuracy of the inter/extrapolation. The maximum deviation of all resonators is \SI{11.3}{MHz} from the mean deviation. The internal quality factor $Q_i$ slightly decreased from \num{2.1e4} to \num{1.8e4} after trimming and the coupling quality factor $Q_c$ stayed unchanged. This decrease is due to the fact that the array was placed at a better position with respect to the focal surface. This fact is actually the only explanation for the higher response measured when we mapped the trimmed array.
\begin{figure}
\includegraphics[width=3.2in,trim={7.7cm 8.3cm 5.8cm 7.5cm}, clip]{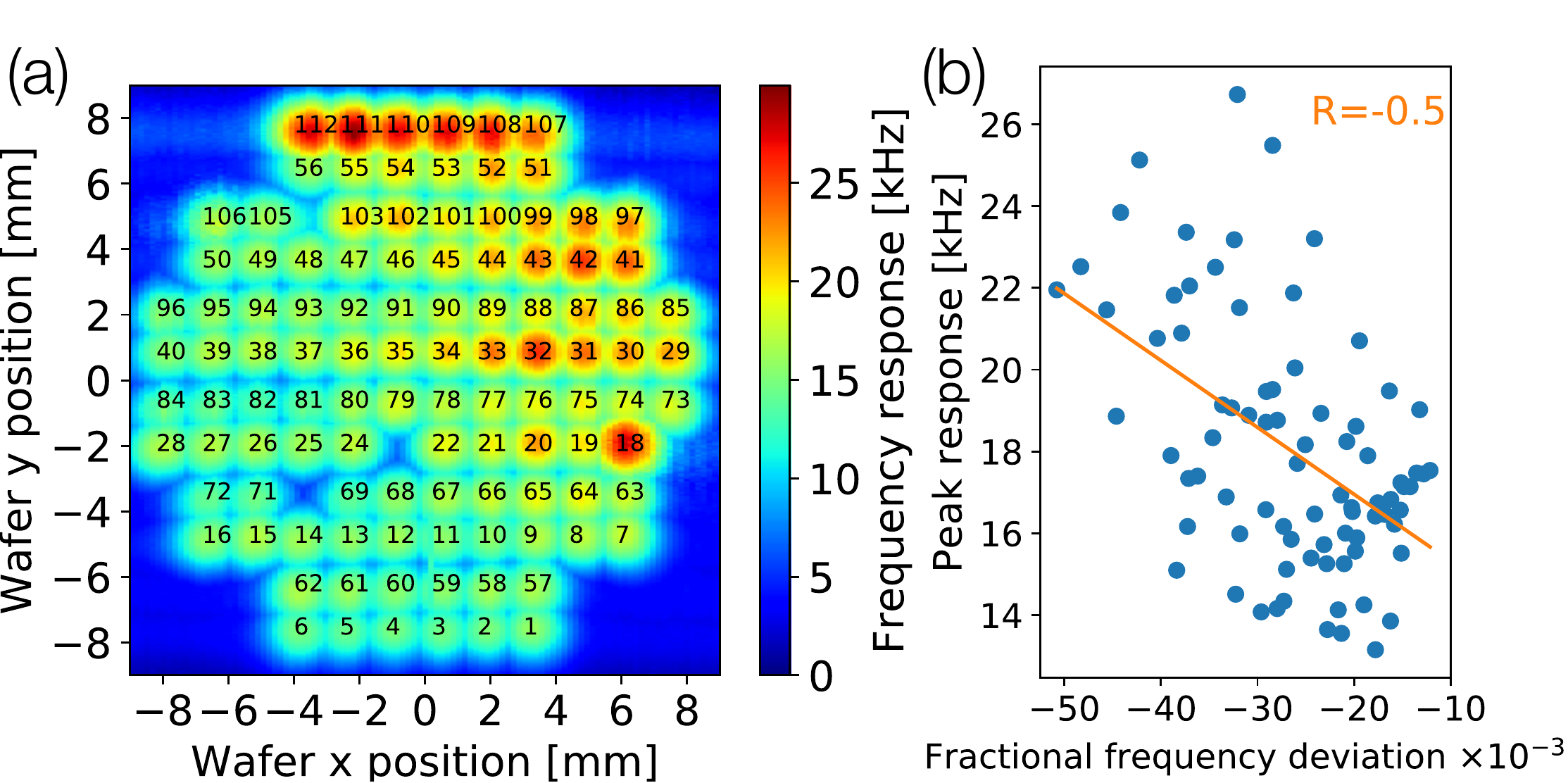}
\caption{\label{fig:beam} (a) Beam mapping of all measured pixels after trimming. $108$ out of $112$ pixels are mapped using the beam mapping system. The resonator indexes in the initial design are shown on each beam. (b) The peak response of each pixel is plotted against its fractional frequency deviation. From the data, a moderate correlation ($R=-0.5$) is found, which partly explains the variation of the optical response.}
\end{figure}

The optical beam mapping, shown in Fig.~\ref{fig:beam}, was measured using the beam mapping system. $108$ out of $112$ pixels ($96.7\%$) are measured. No pixel-to-pixel crosstalk is observed down to the noise level (\SI{-15}{dB}). This confirms that we have eliminated the problem of frequency collision. This improvement is obtained by a reduction of the frequency span, making it smaller than the bandwidth, and an increase of the homogeneity of the resonance frequency spacing. The missing $4$ pixels were identified in the optical microscope and showed fabrication defects. Only one of these pixels was damaged during the trimming fabrication step. The variation of the response of each pixel has a moderate correlation ($R=-0.5$) with the resonance frequency deviation, which is mainly caused by the variation of the inductor width. The responsivity also depends on the quality factor and absorption efficiency by impedance matching, but both these two quantities depend on the inductor width, which makes it hard to do a thorough analysis. The noise around the readout frequency (\SI{23}{Hz}) stays the same after trimming.

The multiplexing factor and the array optical yield is increased from $79$ to $108$ and from $70.5\%$ to $96.7\%$, respectively. 
A minimal resonance frequency spacing \SI{1.66}{\MHz} is calculated using the 5-linewidth frequency collision criterion $5f/Q$, allowing $300$ resonators to be placed within a \SI{500}{\MHz} readout bandwidth, while the design in this Letter only holds 112 resonators. The current NIKA2 1mm array has $1140$ LEKID pixels~\cite{Adam:2018a} on $8$ feedlines. Using the trimming technique described in this Letter, this number could be increased to about $2500$ pixels with high yield.

See supplementary material for the detailed optical setup and the mapping system.

The authors thank F. L\'{e}vy-Bertrand, D. Billon-Pierron and A. Barbier for experimental help, and K.F. Schuster for useful discussions. This work has been partially funded by LabEx FOCUS ANR-1-LABX-0013.

\bibliography{aipsamp}

\end{document}